\documentclass[]{spie}  %>>> use for US letter paper

\usepackage{textcomp, gensymb} % adds "degree"
\usepackage{amsmath,amsfonts,amssymb}
\usepackage{graphicx}
\usepackage[colorlinks=true, allcolors=blue]{hyperref}

\newcommand\farcsec{\mbox{$.\mkern-4mu^{\prime\prime\mkern-2mu}$}}
\newcommand\arcmin{\mbox{$^\prime$}}
\newcommand\arcsec{\mbox{$^{\prime\prime}$}}
\newcommand\electron{\mbox{$\mathrm{e^-}$}}
\makeatletter
\newcommand*{\rom}[1]{\expandafter\@slowromancap\romannumeral #1@}
\newcommand\ion[2]{#1$\;${%
\ifx\@currsize\normalsize\small \else
\ifx\@currsize\small\footnotesize \else
\ifx\@currsize\footnotesize\scriptsize \else
\ifx\@currsize\scriptsize\tiny \else
\ifx\@currsize\large\normalsize \else
\ifx\@currsize\Large\large
\fi\fi\fi\fi\fi\fi
\rmfamily\rom{#2}}\relax}% 
\makeatother

\newcommand\fNL{f_{\rm NL,local}}
\newcommand\Ha{H$\alpha$}
\newcommand\Hb{H$\beta$}
\newcommand\Oii{[\ion{O}{2}]}
\newcommand\Oiii{[\ion{O}{3}]}
\newcommand\Nii{[\ion{N}{2}]}
\newcommand\reg{\textsuperscript{\tiny\textregistered}}
\newcommand\tm{\textsuperscript{\tiny TM}}

\usepackage{multicol}
\makeatletter
\renewcommand\@author{
    \AB@authlist\\[\affilsep]
    \begin{multicols}{2}
        \raggedright
      \begin{quote}
      \small
        \AB@affillist
      \end{quote}
    \end{multicols}
    }
\makeatother

\title{The SIRMOS mission concept: a space-based near-IR spectroscopic survey of 100 million galaxies in the Euclid Wide Survey footprint }

\author{Gregory D. Wirth}
\affil{BAE Systems -- Space \& Mission Systems, Broomfield, CO, USA}

\author{Yun Wang}
\affil{Caltech/IPAC, Pasadena, CA, USA}

\author{Massimo Robberto}
\affil{STScI, Baltimore, MD, USA}

\author{Mark Dickinson}
\affil{NSF NOIRLab, Tucson, AZ, USA}

\author[2]{Thomas Greene}

\author{Jeffrey Kruk}
\affil{NASA GSFC, Greenbelt, MD, USA}

\author{Will Percival}
\affil{Waterloo Centre for Astrophysics, Univ. of Waterloo, Perimeter Institute for Theoretical Astrophysics, Canada}

\author[5]{Joshua E. Schlieder}

\author[2]{Andreas Faisst}

\author{Lynne Hillenbrand}
\affil{Caltech, Pasadena, USA}

\author{Jeyhan Kartaltepe}
\affil{Rochester Institute of Technology Laboratory for Multiwavelength Astrophysics, Rochester, NY, USA}

\author{Nikhil Padmanabhan}
\affil{Yale Univ., New Haven, CT, USA}

\author{Lado Samushia}
\affil{Kansas State Univ., Manhattan, KS, USA}

\author{Robert Content}
\affil{AAO, Macquarie Univ., Sydney, Australia}

\author{Micol Bolzonella}
\affil{INAF OAS, Bologna, Italy}

\author{Samuel Brieden}
\affil{Institute for Theoretical Particle Physics and Cosmology (TTK), RWTH Aachen, Germany}

\author[2]{Lee Armus}

\author{Benedict Bahr-Kalus}
\affil{INAF Osservatorio Astrofisico di Torino, Italy}

\author{Andrew Benson}
\affil{Carnegie Observatories, Pasadena, CA, USA}

\author{Florian Beutler}
\affil{Univ. of Edinburgh, Scotland}

\author{Jarle Brinchmann}
\affil{Instituto de Astrofísica e Ciências do Espaço, Univ. do Porto, Portugal and
European Southern Observatory, Garching, Germany}

\author[1]{Hansford Cutlip}

\author{Emanuele Daddi}
\affil{AIM - Université Paris-Saclay, CEA, CNRS, France}

\author[2]{Kyle Finner}

\author[1]{Jason Forsyth}
\author[1]{David Glaister}
\author[1]{Carlita Gorham}

\author{Andrew Hearin}
\affil{Argonne National Laboratory, Lemont, IL, USA}

\author{Cullan Howlett}
\affil{School of Mathematics and Physics, Univ. of Queensland,  Australia}

\author[11]{Jon Lawrence}

\author[5]{Gregory Mosby}

\author{Eva-Maria Mueller}
\affil{Univ. of Sussex, U.K.}

\author{Seshadri Nadathur}
\affil{Univ. of Portsmouth, U.K.}

\author[2]{Kalina V. Nedkova}

\author[8]{Zoran Ninkov}

\author{Ken Osato}
\affil{Chiba Univ., Japan}

\author{Casey Papovich}
\affil{Texas A\&M Univ., College Station, TX, USA}

\author[14]{Jack Piotrowski}

\author[12]{Lucia Pozzetti}

\author{Alvise Raccanelli}
\affil{Univ. of Padova, Italy}

\author{Jason Rhodes}
\affil{JPL, Pasadena, CA, USA}

\author{Shun Saito}
\affil{Missouri S. \& T. Univ., Rolla, MO, USA}

\author[1]{Rebecca Schindhelm}

\author{Hee-Jong Seo}
\affil{Ohio Univ., Athena, OH, USA}

\author{Zachary Slepian}
\affil{Univ. of Florida, Gainesville, FL, USA}

\author{Steve Smee}
\affil{Johns Hopkins Univ., Baltimore, MD, USA}

\author[2]{Francesco Spezzati} 

\author[11]{Lee Spitler}

\author[1]{Scott Streetman}

\authorinfo{Further author information: (Send correspondence to G.D.W.)\\G.D.W.: E-mail: gregory.wirth@baesystems.us, Telephone: +1 303 939 7401}

\begin{document} 
\maketitle
\begin{abstract}
SIRMOS is a mission concept for a future NASA Astrophysics Explorer
opportunity that aims to revolutionize our understanding of the
universe by mapping the large-scale structure of galaxies in
unprecedented breadth and detail to explore cosmic origins and
evolution. Specifically, the primary science goals of SIRMOS are to
discover how cosmic inflation seeded large-scale structure, reveal the
nature of the accelerating expansion of the universe, and measure the
sum of neutrino masses to exceptionally high precision. Its secondary
science goals include determining how environment shapes galaxy
evolution at cosmic noon and beyond, and probing the Milky Way's
dust-enshrouded regions, reaching the far side of our Galaxy.

To accomplish its transformative science goals, SIRMOS executes a
3-year cosmology survey over the \textit{Euclid} footprint of
$\sim$14,000~deg$^2$, obtaining slit spectra for $\sim$100 million galaxies
at $1<z<4$. SIRMOS is the first space mission designed specifically to
leverage \textit{Euclid}'s wide near-infrared (IR) imaging survey for target
selection and employs a digital micro-mirror device (DMD) to acquire
up to 4,000 simultaneous spectra over a wide $0.9~\mathrm{deg}^2$ field of
view. The instrument's excellent spectral resolution of $R=1300$ and
continuous wavelength coverage from 1.25--2.5~$\mu$m capture the
redshifts of distant galaxies with no gaps, reducing systematic
biases to achieve ground-breaking science.
\end{abstract}


\section{INTRODUCTION}
\label{sec:intro}  % \label{} allows reference to this section

A generation ago, the emergence of dedicated, highly-multiplexed
spectroscopic instruments on 8--10~m class telescopes enabled a golden
age of mapping the large-scale three-dimensional distribution of
galaxies. Wide-field surveys like 2dFGS covered large areas of the sky to a moderate 
depth while ``pencil beam'' surveys like DEEP and DEEP2 on Keck probed far 
deeper but over more restricted areas of the sky, ultimately
delivering new insights into both cosmology and galaxy 
evolution.\cite{2005MNRAS.362..505C,Weiner_2005,2006ApJ...638..668C} 
More recently, the Dark Energy Spectroscopic Instrument (DESI)
on the KPNO 4-m telescope has gathered redshifts of over 10 million
galaxies and a million quasars, demonstrating that a highly-efficient
multi-object spectrograph (MOS) can generate rich datasets even when
coupled with telescopes of relatively modest aperture.\cite{2025PhRvD.112h3511L,2025PhRvD.112h3513E}

\subsection{The importance of an IR MOS in space}

The groundbreaking surveys cited above employed spectrographs
operating in the visible part of the electromagnetic spectrum; 
consequently, they are optimized to measure galaxies in the redshift range
of $0.3 \lesssim z \lesssim 1.5$ where ground-based telescopes can
capture the key diagnostic emission lines used for measuring redshifts (\Ha~$\lambda$6563~\AA,
\Hb~$\lambda$4861~\AA, 
\Oii~$\lambda\lambda$3726,3729~\AA,
\Oiii~$\lambda\lambda$4959,5007~\AA, 
and 
\Nii~$\lambda$6583~\AA). 

We can achieve more powerful probes of key cosmological parameters by
observing more distant galaxies at redshifts out to $z\approx4$, but
those key spectral features shift out of the visible and into the
near-infrared. While infrared multi-object spectrographs have now been
deployed on major telescopes like Keck (MOSFIRE\cite{10.1117/12.924794}) and VLT (KMOS\cite{10.1117/12.2055496}), the
quality of data gathered by these instruments is limited by the
blurring effect of the Earth's atmosphere, the opacity of the 
atmosphere at wavelengths where water and other 
species absorb significant light, and the high background 
generated by thermal and telluric emission --- OH airglow 
lines in particular ---  that contribute substantial noise 
and limit the ability to observe faint targets. The opacity 
gaps due to water vapor also produce broad gaps in redshift 
coverage for key spectral lines in  the galaxy spectra.
Ultimately this leads to redshift-dependent biases in the
data which are challenging to model. These redshift-dependent
selection effects contribute uncertainty to the derived cosmological
parameters.

Putting an IR MOS into orbit confers important advantages over similar
instruments on the ground. 
The absence of atmospheric absorption and emission lines and the cooler instrument temperatures
achievable in space both reduce the noise resulting from the IR night
sky and telescope emission and eliminate the problematic gaps in
redshift coverage, reducing bias in measurements of the galaxy
distribution. Observing above the atmosphere also reduces 
time-varying atmospheric blur typical of ground-based telescopes that
contributes to source confusion and degrades the signal-to-noise ratio ($S/N$). 
The NIRSpec instrument on JWST is one such MOS
instrument that can exploit these advantages of observing from
space. However, the scarcity of JWST observing time, the relatively
low multiplexing capability of NIRSpec, the relatively small field
of view (FoV), and the long slew and settle time for JWST all 
conspire to limit the effectiveness of JWST
as a survey instrument on large scales. An IR space telescope of
more modest aperture but large field of view coupled with a highly capable IR MOS is well
positioned to make major strides in mapping the Universe in the
critical redshift range $1<z<4$ beyond the grasp of its ground-based
counterparts.

This paper describes our concept for such a mission: the Satellite for
Infrared Multi-Object Spectroscopy (SIRMOS). This mission concept
draws from earlier incarnations in the form of the ATLAS Probe mission
\cite{Wang_2019, 10.1117/12.2313082} and was previously conceived as a
``small explorer'' (SMEX) mission \cite{10.1117/12.3017865} before being recast as a
more capable spacecraft planned for submission to the next NASA
astrophysics ``medium explorer'' (MIDEX) opportunity.

\subsection{Complementarity with existing and planned facilities}

\begin{figure}[tbp]
    \centering
    \includegraphics[width=0.5\linewidth]{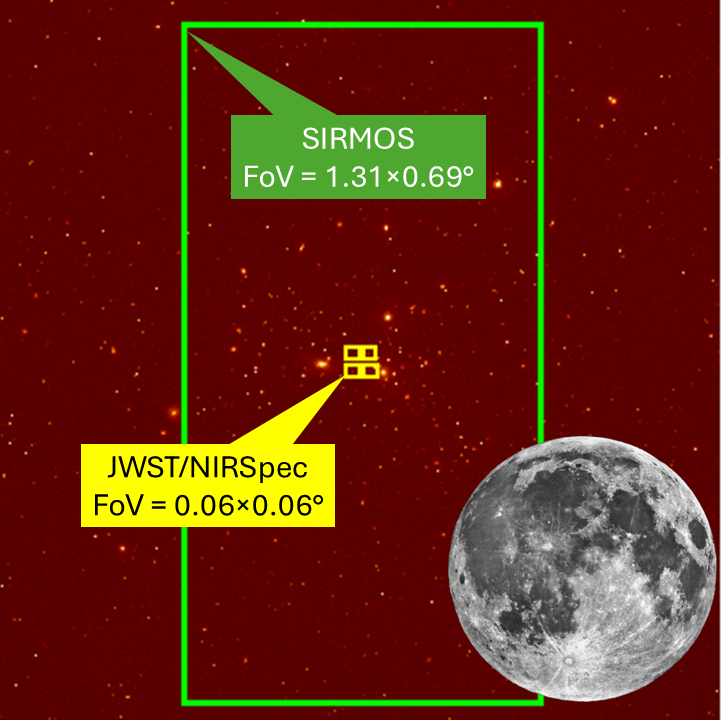}
    \caption{\label{fig:FOV}
The SIRMOS field of view is 350 times larger than that of
  JWST's NIRSpec in its multi-object spectroscopy mode, making it the superior option for a wide-field redshift survey. Full moon shown for scale.} 
\end{figure}

SIRMOS features superior multi-spectral capabilities compared to
other facilities in orbit. During its operational lifetime, which ended
in 2008, \textit{Hubble}'s NICMOS instrument offered near-IR slitless multi-spectral
capability; however, this observing mode suffers from low spectral
resolution, potential source confusion, and elevated noise due to the warm
telescope optics and elevated background light inherent in slitless
modes.\cite{Thompson_1998} JWST's more powerful NIRSpec instrument is equipped
with micro-shutter arrays capable of true multi-slit spectral operation,
achieving a multiplex factor of $\sim100$ with spectral resolution
options of $R=100$, 1000, and 2700 coupled to a large-aperture
IR-optimized telescope, making this the instrument of choice for
targeting the faintest and most distant galaxies in the early 
Universe.\cite{10.1117/12.2231837} However, the relatively 
small $3\arcmin\times3\arcmin$ NIRSpec field of view and the 
telescope's lengthy slew-and-settle times severely limit its 
capabilities as a large-scale survey instrument.  Figure~\ref{fig:FOV}
compares the relative fields of view for JWST's NIRSpec and the SIRMOS
instrument.

The pending \textit{Roman} and operational \textit{Euclid} 
spacecraft have smaller apertures but much wider
fields of view compared to JWST, enabling surveys of vast portions of
the sky to capture broadband colors and slitless spectra of modest
resolution for tens of millions of galaxies.  The resulting
spectra posess sufficient nominal spectral resolution to measure
moderately accurate galaxy redshifts, but because they observe in
slitless mode the resulting data are subject to confusion from
closely spaced galaxies and reduced spectral resolution due to
the finite spatial extent of galaxies.  SIRMOS achieves similar
spectral resolving power but its slitmask-based design reduces
source confusion and produces consistent spectral resolution from
target to target, yielding superior redshift accuracy.

The recently launched SPHEREx spacecraft \cite{Bock2026} is currently conducting a
revolutionary all-sky survey of targets with a powerful instrument
that observes the sky in 96 color bands over the 0.75--5.0~$\mu$m
wavelength range, capturing spectral classifications and redshifts for
up to 450 million galaxies.\cite{10.1117/12.2312860} 
In the 1--2.5~$\mu$m wavelength
range, SPHEREx achieves a resolving power of $R=41$ compared to the
$R=1300$ of SIRMOS, making the latter a superior facility for 
mapping the three-dimensional galaxy distribution in detail. 
Nevertheless, SPHEREx will generate a massive redshift database classifying $\sim$16 million galaxies at 0.3\% redshift uncertainty and another $\sim$470 million galaxies at better than 10\% precision.

\begin{figure}[tbp]
    \centering
    \includegraphics[width=0.9\linewidth]{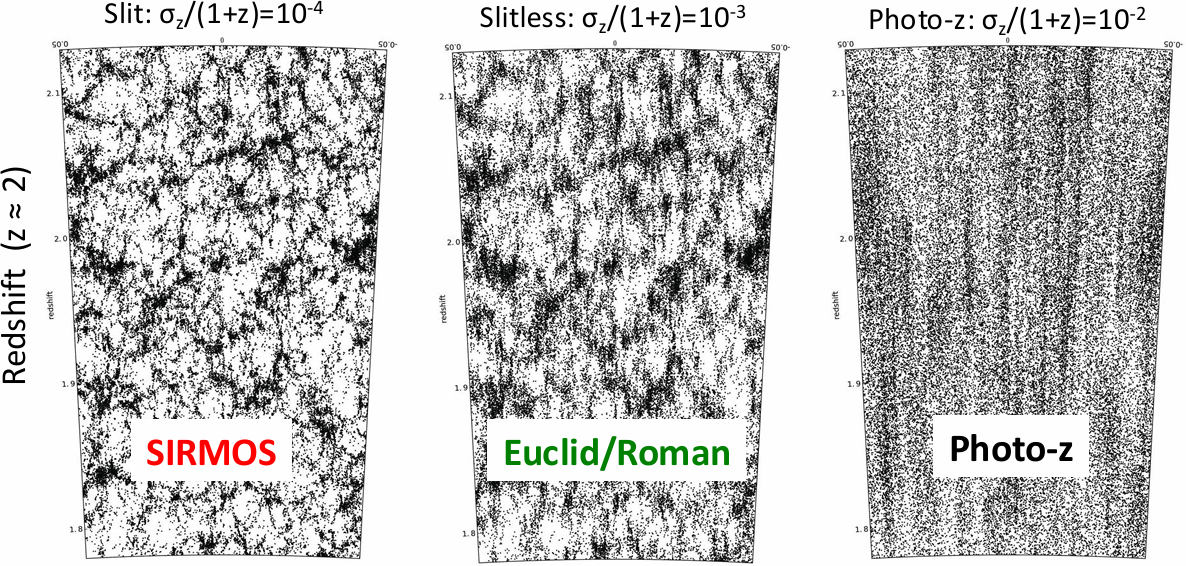}
\caption{Comparison of simulated redshift maps obtained by SIRMOS (left),
\textit{Euclid/Roman} (center), and a multi-band photometric survey (right), assuming their predicted
spectral resolution. Compared with the poorer resolution of the lower-precision surveys that washes out cosmic walls and voids, SIRMOS resolves fine structures in the large-scale
distribution of galaxies, yielding improved estimates of cosmological parameters. \textit{(Figure courtesy of Alvaro Orsi)}}
\label{fig:redshiftAccuracy}
\end{figure}

Figure~\ref{fig:redshiftAccuracy}
shows the relative redshift precision of a moderate-resolution spectral survey like SIRMOS compared to the lower-resolution slitless spectroscopy produced by \textit{Euclid} or \textit{Roman}, and a hypothetical multi-band photometric survey with 1\% redshift uncertainty. The comparison illustrates the power of high-quality spectra from SIRMOS to accurately map details in the galaxy distribution.

SIRMOS builds on the SPHEREx heritage to achieve its science objectives. Coupling an expansive FoV and superior multiplexing
capability with excellent spectral resolution allows SIRMOS to address
fundamental questions in cosmology. As described in \S\ref{sxn:science} below, SIRMOS 
explores the physics of cosmological inflation, addresses the
acceleration of the cosmic expansion rate, and measures the sum of
neutrino masses to unparalleled precision. Each of these goals
requires mapping the large-scale distribution of distant galaxies via
a dedicated high-accuracy galaxy redshift survey that illuminates the
origin and evolution of the Universe.

The CASTOR (Cosmological Advanced Survey Telescope for Optical and ultraviolet Research) mission under conceptual development in Canada shares the objective of deploying a DMD-based multi-object spectrometer for wide-field surveys.\cite{10.1117/1.JATIS.11.4.042202}  Similar to SIRMOS, CASTOR provides the spectral resolution necessary to resolve key diagnostic lines.  However, the two missions operate in distinct wavelength regimes and target different scientific objectives; while the CASTOR MOS is optimized for investigating active galactic nuclei and blue compact dwarfs in the local universe, SIRMOS is designed to address fundamental cosmological questions via a wide-field infrared redshift survey. As a result, the two missions are fully complementary.

SIRMOS P.I. Yun Wang (Caltech) has championed the mission concept by
assembling an international team of scientific collaborators drawn from
academic institutions in the U.S., U.K., Australia, Canada, Italy, France,
Germany, Japan, and the Netherlands. The SIRMOS team includes optics
design expertise from AAO, hardware expertise from GSFC, STScI, and
RIT, and is partnered with BAE Systems Space \& Mission Systems to
implement the scientific vision. 
 
\section{Design}

\begin{figure}[tbp]
    \centering
    \includegraphics[width=0.32\linewidth]{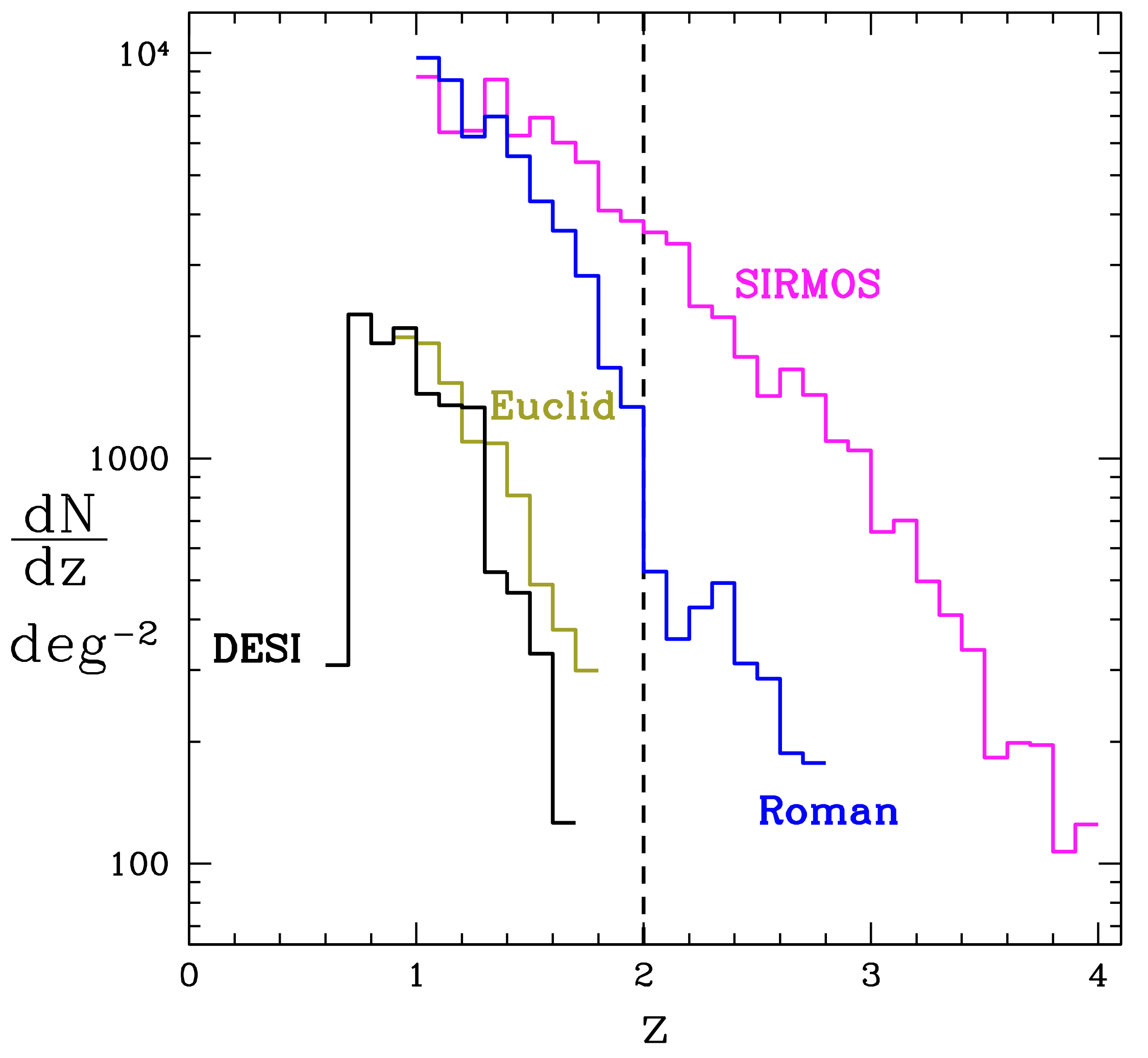}
    \includegraphics[width=0.32\linewidth]{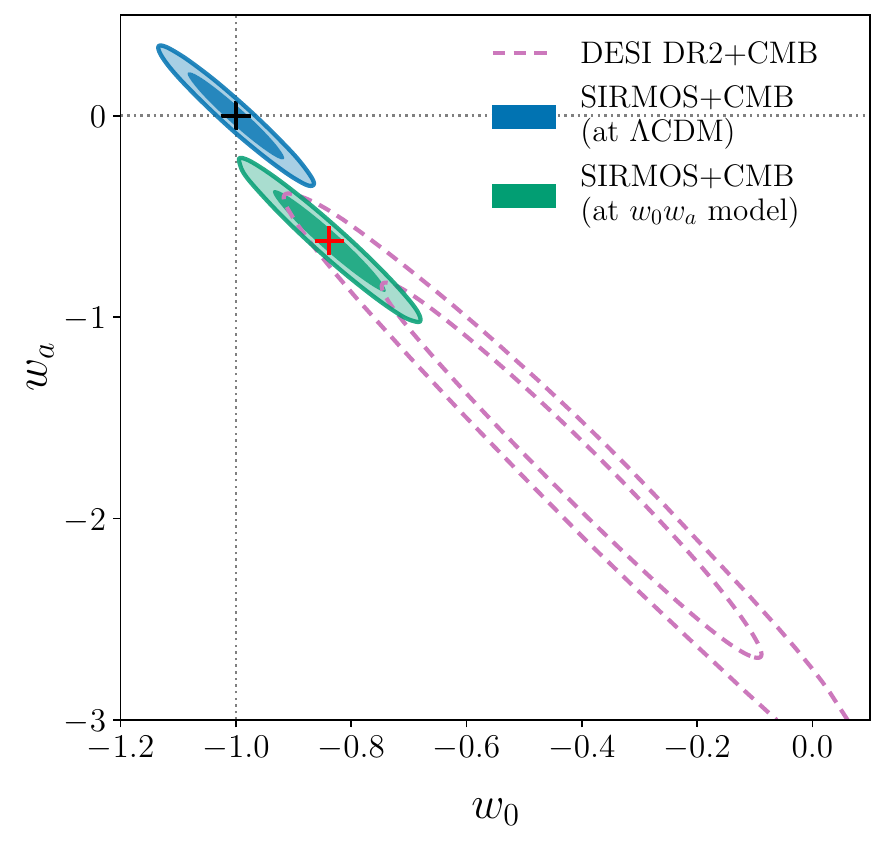}
    \includegraphics[width=0.32\linewidth]{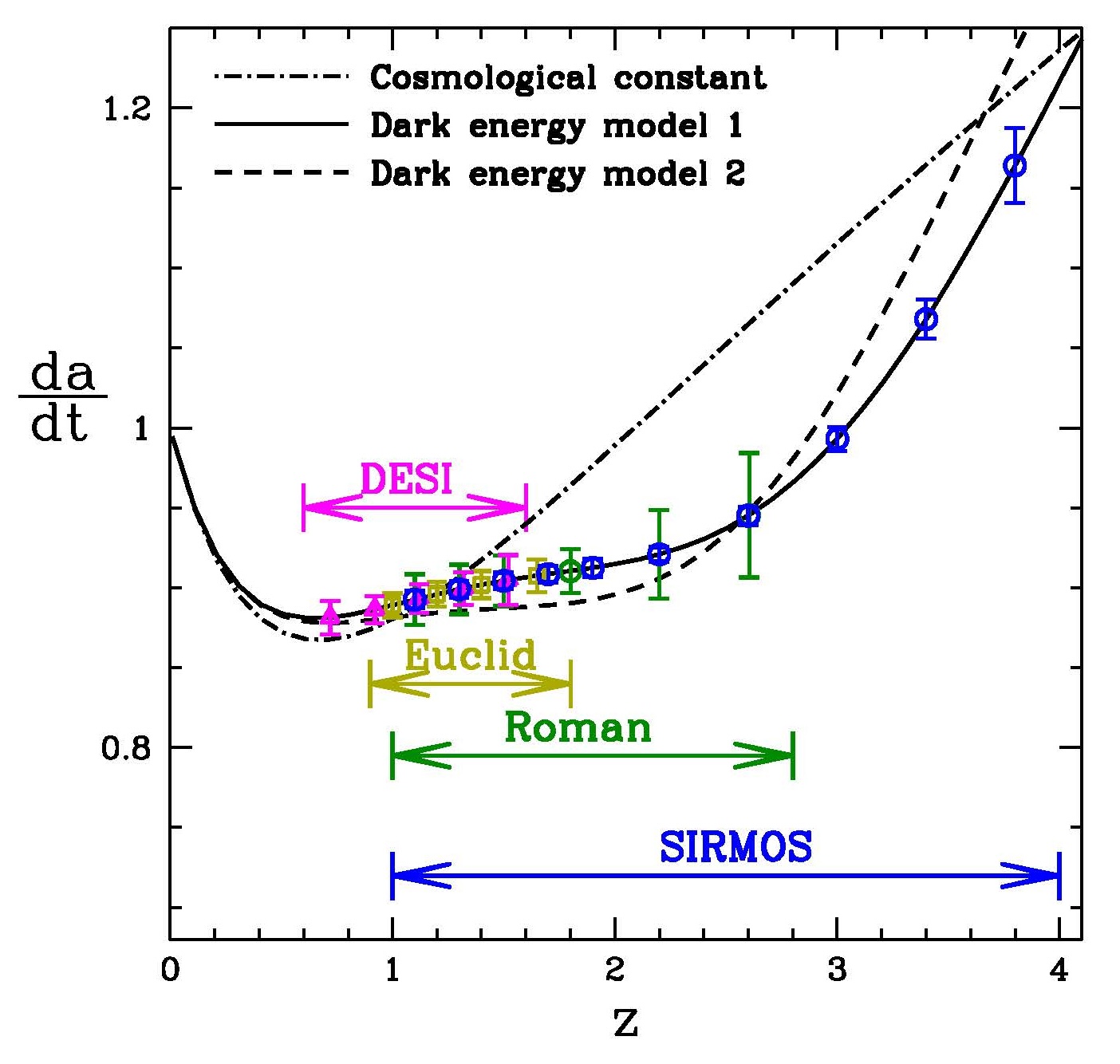}%\hspace{0.4in} 
   \caption{\textit{Left:}~Comparison of detectable galaxy number
     density for different spectroscopic surveys. \textit{Center:}~Predicted
     constraints on dark energy parameters from SIRMOS in two
     scenarios compared to recent results from DESI
     survey. \textit{Right:}~Time derivative of cosmic scale factor
     $a$ vs. redshift $z$ compared to predicted measurement
     uncertainties for various surveys, illustrating how the greater
     redshift sampling and lower measurement uncertainty of SIRMOS
     distinguishes dark energy models with more complex high-$z$
     behavior.  }
   \label{fig:dadt}
\end{figure}

Achieving the ambitious science objectives described more fully below in \S\ref{sxn:science} requires an understanding of the detailed three-dimensional distribution of galaxies over a massive volume. Such knowledge requires a extensive catalog of high quality galaxy redshifts on large scales plus careful control and knowledge of systematic sources of error. No existing or planned facility has the combination of features capable of generating this dataset (see Fig.~\ref{fig:dadt}).

Meeting the SIRMOS science goals is possible only under demanding observing conditions: high-precision galaxy redshifts collected across the entire spectral range of interest free of wavelength gaps, stable backgrounds, and measurement calibrations that remain consistent over time.  Surveys conducted from the ground suffer from atmospheric absorption and OH emission line contamination that is both extensive and time varying, introducing significant and uncontrollable interference with redshift measurements. Near-IR observations must contend with both reduced atmospheric transmission and rapidly evolving backgrounds; these are challenging to model and contribute to measurement error. Placing the instrument in orbit solves all of these problems, enabling continuous monitoring of the 1.25--2.5~$\mu$m wavelength range with background contamination which is both well controlled and stable, allowing calibrations to remain stable and systematic errors to be limited and understood to a high degree so that the redshift measurements are unbiased.

To map the cosmic structure in sufficient depth and detail to achieve its demanding science objectives, SIRMOS must survey galaxies across the full $1<z<4$ redshift range within a massive volume covering 500~Gpc$^3$ over an area of 14,000~deg$^2$. The goal is to collect $10^8$ redshifts over the projected 3-year mission lifetime by identifying the key emission features characteristic of the star-forming galaxies that prevail at earlier epochs in the history of the Universe (\Ha, \Oiii, and \Oii). The resulting deep and uniform map of the large-scale galaxy distribution will feature stringent controls on sources of systematic error, enabling precise measurements of fundamental physical parameters. By employing true multi-slit spectroscopy capability enabled by the Digital Micromirror Device (DMD), SIRMOS measures redshifts accurate to $\sigma_z/(1+z)\simeq10^{-4}$, improving upon the accuracy of redshifts from \textit{Euclid} and \textit{Roman} by an order of magnitude. As described below, the SIRMOS payload and spacecraft employ key technologies from successful missions previously developed by industry partner BAE Systems Space \& Mission Systems.

\subsection{Mission Concept}

SIRMOS is a mission concept aimed at the next NASA MIDEX opportunity. The payload features a 60~cm primary mirror with a wide $0.9~\mathrm{deg}^2$ FoV that images the field onto a Digital Micromirror Device (DMD) serving as a reflective electronic slitmask. The high-throughput, low-resolution spectrometer captures spectra of 4,000 targets at once. The spacecraft bus leverages the successful design of SPHEREx and occupies a sun-synchronous, terminator-aligned low-earth orbit to maintain thermal stability. To collect redshifts to $10^8$ galaxies over the 3-year mission lifetime, SIRMOS integrates for a total of 51~min per field divided into 17~min of exposure time over 3 separate pointings occurring at 12-month intervals. The SIRMOS spacecraft maintains stable instrument pointing during the 17~min of exposure time, which is divided into multiple sub-exposures to enable rejection of cosmic rays and permit integral-pixel dithering. To increase the science return, targets observed with sufficient quality to measure a secure redshift ($S/N\ge5$) are replaced with new targets on subsequent visits while those with lower $S/N$ are re-observed. 

\subsection{Payload Summary}

\begin{figure}[tbp]
    \centering
    \includegraphics[width=0.5\linewidth]{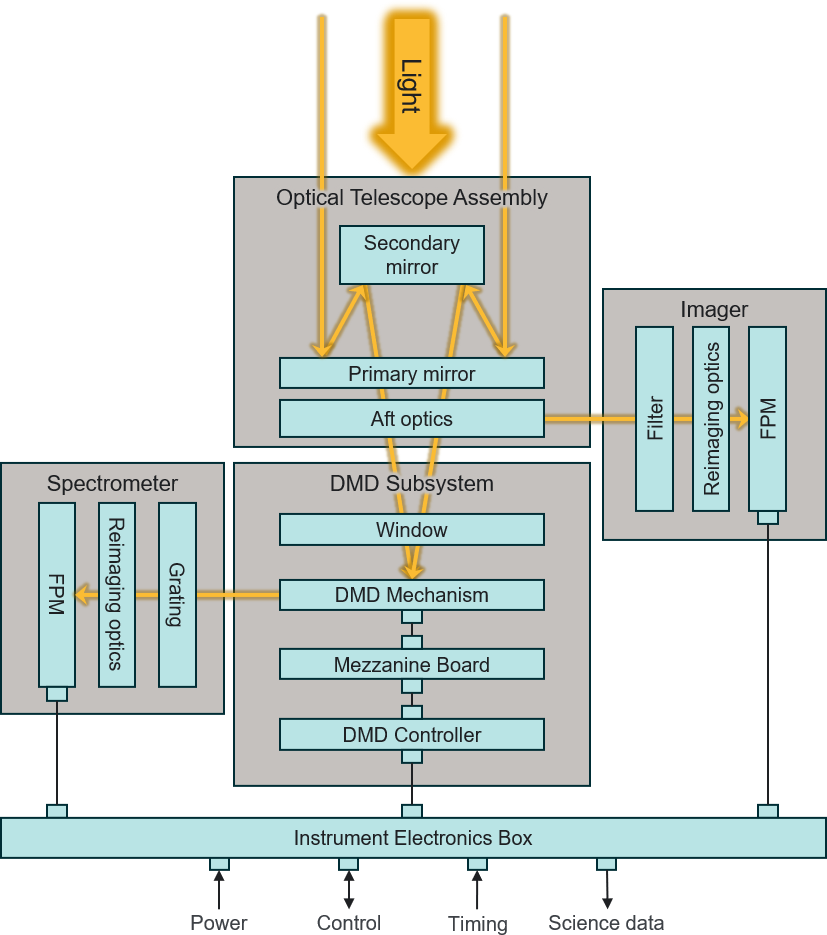}
    \caption{SIRMOS payload Functional Block Diagram}
    \label{fig:fbd}
\end{figure}

To deliver the intended sample of $10^8$ distant galaxy redshifts 
needed to map the large-scale cosmic structure within the survey 
volume over the 3-year prime mission lifetime, SIRMOS features 
a wide field of view with exceptionally flexible multi-object 
spectroscopy capability in a compact package. 
As shown in Fig.~\ref{fig:fbd}, the simple SIRMOS payload design
incorporates an imager and spectrometer within a limited volume
and employs a single mechanism. The telescope and aft optics feed light
to the DMD, which activates selected micro-mirrors to direct light for
specific targets to the spectrometer for redshift measurements. An
off-axis imager views a field adjacent to the spectroscopy field to
monitor on-sky alignment. Both the spectrometer and imager feature
low-noise array detectors for high $S/N$ observations.  Table~\ref{tab:sirmosParameters} lists the key payload
parameters.

\begin{table}[hb]
    \centering
    \includegraphics[width=0.75\linewidth]{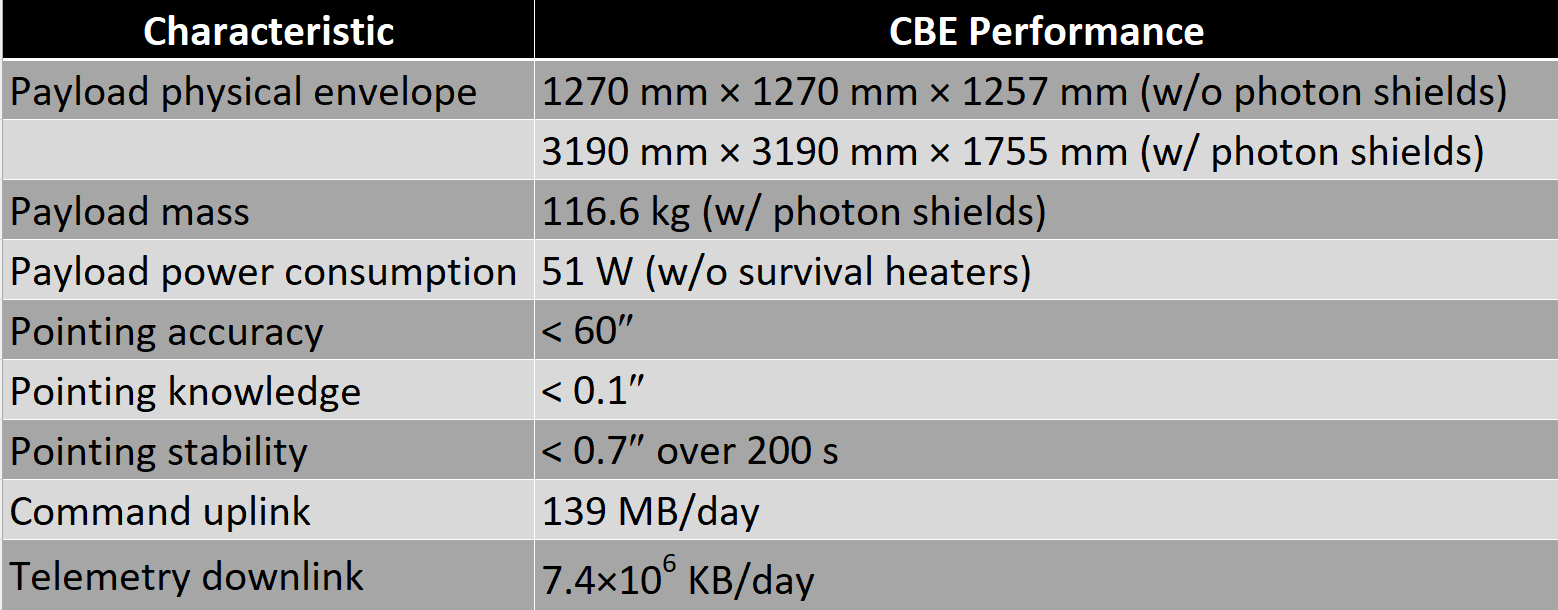}
    \caption{SIRMOS payload key parameters}
    \label{tab:sirmosParameters}
\end{table}

\subsection{Telescope Requirements \& Estimates}
\begin{figure}[tbp]
    \centering
    \includegraphics[width=0.5\linewidth]{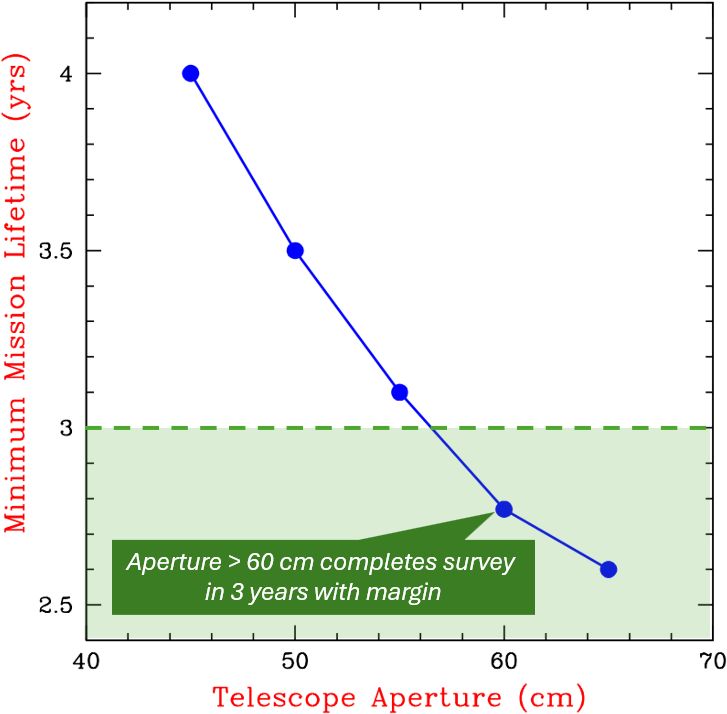}
    \caption{SIRMOS requires a minimum 60~cm diameter primary mirror to capture $10^8$ redshifts in 3 years}
    \label{fig:aperture}
\end{figure}

SIRMOS requires a wide field of view coupled with sufficient light-gathering power to detect faint emission lines within the allocated 51~min of exposure time per field. Table~\ref{tab:sirmosTelescope} lists the baseline telescope parameters.  As shown in Fig.~\ref{fig:aperture}, science requirements impose a lower limit of 60~cm on the telescope aperture size. Payload cost is not strongly dependant on aperture size as long as the total payload mass remains within the limit imposed by the bus.

\begin{table}[h!]
    \centering
    \includegraphics[width=0.6\linewidth]{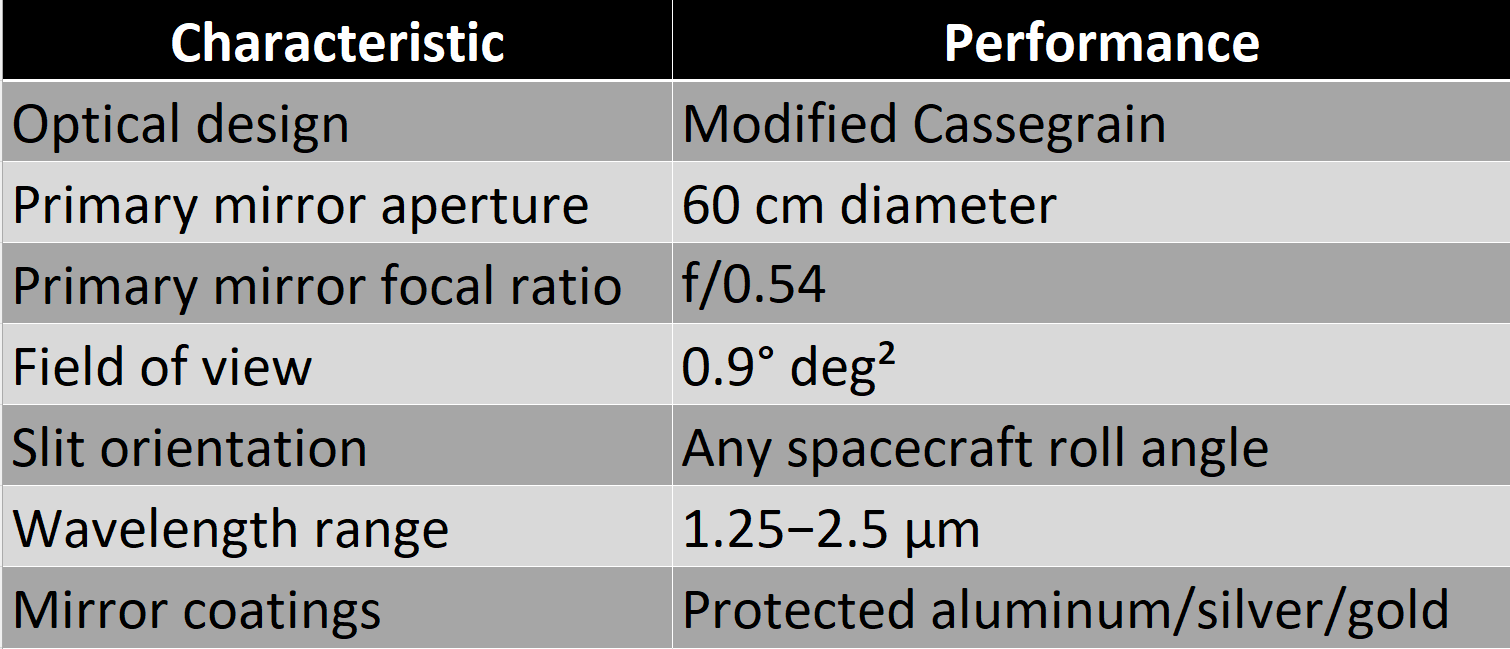}
    \caption{SIRMOS telescope characteristics}
    \label{tab:sirmosTelescope}
\end{table}

\begin{figure}[tbp]
    \centering
    \includegraphics[width=0.5\linewidth]{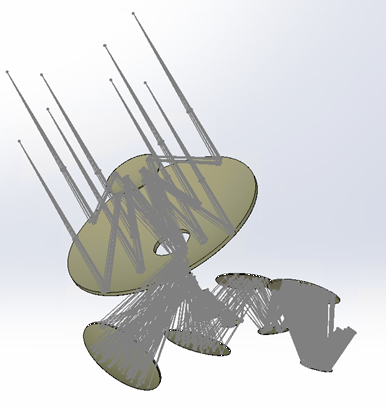}
    \caption{SIRMOS telescope elements deliver wide-field, diffraction limited images to the DMD}
    \label{fig:lightPath}
\end{figure}

The sophisticated SIRMOS optical design (described more fully in  the accompanying poster presented by co-Investigator and optical designer Robert Content at this conference) achieves an ample
$0.9~\mathrm{deg}^2$ field of view by coupling a short focal length primary
mirror with aft optics that fold and focus the beam onto the DMD.\cite{10.1117/12.3017865} Fig.~\ref{fig:lightPath} shows a ray trace of the all-reflective
design which is cooled to 140~K to limit thermal background. The
all-aluminum telescope design features numerous aspheric surfaces
which preserve image quality; future iterations of the optical design
will consider manufacturability and ease of integration as key
constraints.

The absolute pointing requirements listed in Table~\ref{tab:sirmosParameters} are reasonable and
achievable using conventional technology BAE-SMS has already deployed
on other missions. Ensuring that targets fall within the
$2\farcsec3\times2\farcsec3$ area of the DMD micromirrors to which
they are assigned requires an ultimate absolute pointing accuracy
$<1\farcsec0$ in $x$ and $y$ and 10\arcsec\ in roll. SIRMOS achieves
accurate pointing in a 2-stage process. First, bus-mounted star
trackers provide initial coarse alignment (tolerance of 60\arcsec) to
roughly center the targets within the instrument FoV. Second, the
on-board processor uses the imager to autonomously achieve fine
alignment to sub-arcsec precision by measuring the observed pixel
locations of multiple alignment stars specific to the field (uploaded
from the \textit{Euclid} catalog on the ground) and adjusting the
spacecraft tip, tilt, and roll until these measured positions coincide
with the predicted positions. This MOS alignment method is customary
on ground-based telescopes and straightforward to implement on
orbit.\cite{10.1117/12.926693}

The SIRMOS pointing stability requirements are also well defined and
achievable.  SIRMOS keeps 90\% of light on the DMD micromirror (which
serves as the effective spectrometer slit) to achieve the desired
$S/N$. This translates to a $1\sigma$ pointing stability of $<
0\farcsec72$ over the 200~s duration of a SIRMOS spectral
exposure. This is within the on-sky pointing stability performance of
the similar BAE-SMS-built SPHEREx spacecraft which delivers $1\sigma$
stability of $<0\farcsec45$ over timescales of 150~s. SIRMOS uses
similar attitude control components to achieve comparable pointing
performance.

\subsubsection{Spectrometer Requirements \& Estimates }

The SIRMOS spectrometer is both powerful and simple,
inspired by the SAMOS DMD-based spectrograph
developed by SIRMOS team members and successfully deployed on the 4.1-m
SOAR telescope in Chile.\cite{10.1117/12.2311876} The SIRMOS
spectrometer consists of a target selection device, dispersing
element, re-imaging optics, and a focal plane module. See
Table~\ref{tab:sirmosSpectrometer} for spectrometer characteristics.

\begin{table}[h!]
    \centering
    \includegraphics[width=0.75\linewidth]{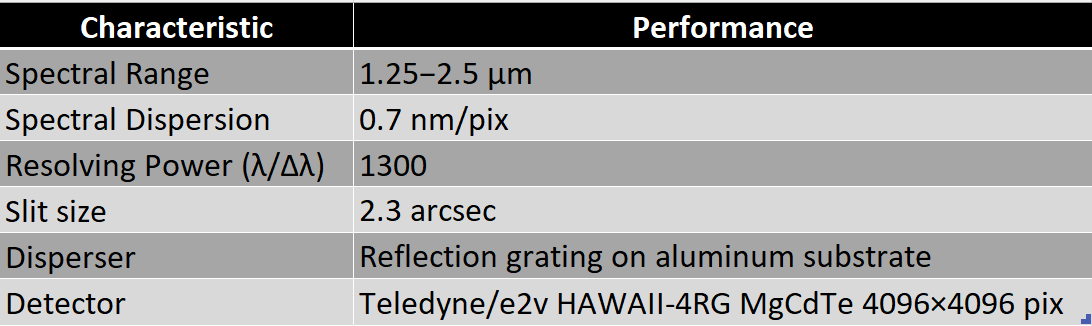}
    \caption{SIRMOS spectrometer characteristics}
    \label{tab:sirmosSpectrometer}
\end{table}

\begin{figure}[tbp]
    \centering
    \includegraphics[width=0.8\linewidth]{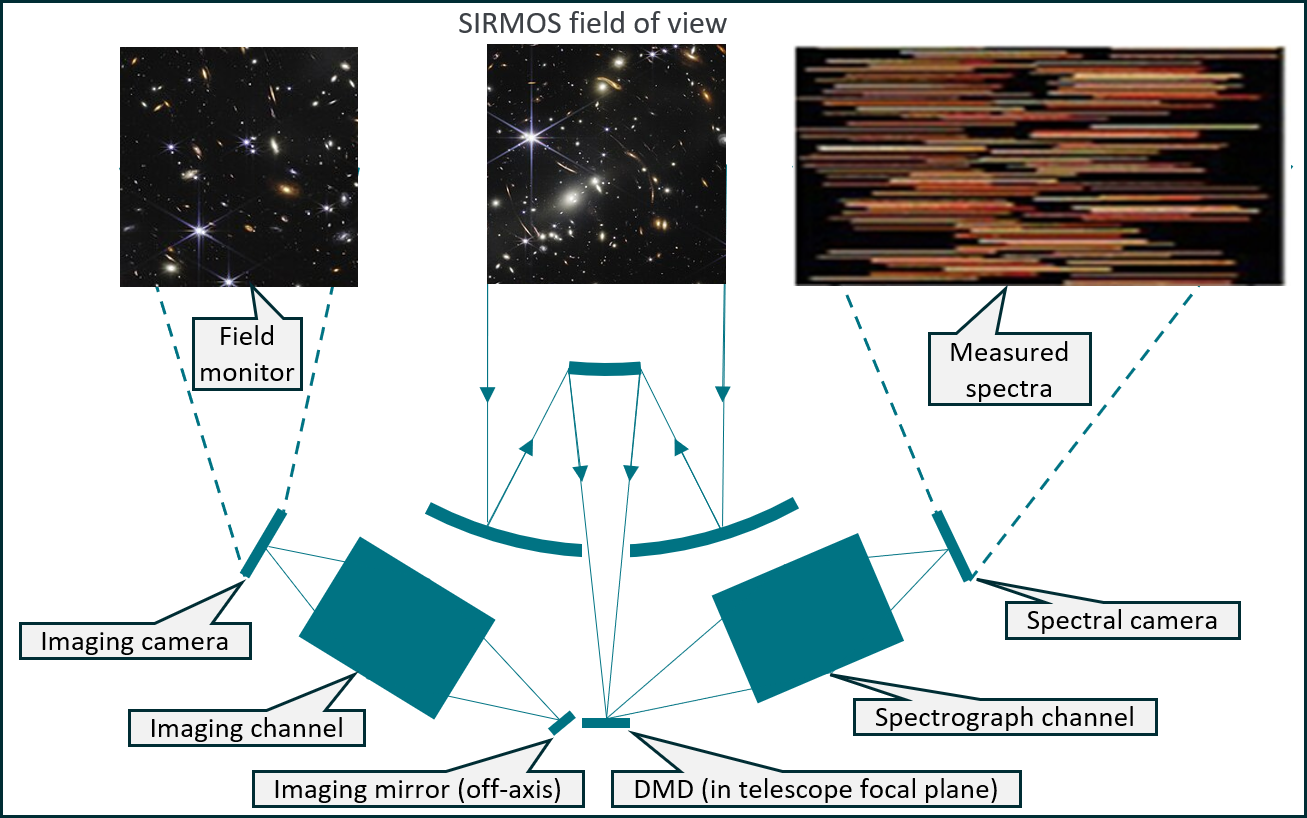}
    \caption{DMD directs light from up to 4000 selected targets to the
      spectral channel for highly multiplexed operation. } 
    \label{fig:sirmosSchematic}
\end{figure}

Rather than a conventional slitmask\cite{10.1117/12.460346} 
used in ground-based telescopes, SIRMOS achieves exceptional target
multiplexing by using a DMD to direct light from selected targets in
the field into the spectrometer.\cite{10.1117/1.JATIS.11.4.042208} The
baseline device is the proven and highly reliable CINEMA\reg~2K
$2048\times1080$ DMD from Texas Instruments; its combination of large
(13.7~$\mu$m) micromirror size and number 
of micromirrors is optimal for astronomical
applications.\cite{10.1117/12.809542} Based on information uploaded
from the ground for the specific field, the on-board DMD controller
activates selected micromirrors corresponding to the telescope focal
plane positions of the observed targets. Light from these micromirrors
reflects off the collimator, onto the grating, and into the spectrometer camera, while
non-activated DMD elements reflect light in a different direction
(``light dump'') that does not enter the
spectrograph. Figure~\ref{fig:sirmosSchematic} illustrates how the
SIRMOS payload employs the DMD for spectroscopy.

The CINEMA\reg~2K DMD is at an appropriate level of technological
readiness for the conceptual stage of SIRMOS. The baseline device was
originally considered by ESA for the \textit{Euclid} mission and
successfully passed a suite of space qualification tests. On the
ground it has been operating reliably for years both for commercial
applications (as the leading technology for digital cinema projectors)
and for astronomy (the ground-based SAMOS instrument on the 4-m
SOAR telescope being the latest example).\cite{10.1117/12.3020796} 
Boulder-based LASP will fly
a DMD on a NASA-sponsored rocket experiment in 2026 to establish
compatibility with the space
environment.\cite{10.1117/12.2594651,2025AAS...24614105P} The
remaining task is to demonstrate operation in the near-IR; DMDs have
been successfully tested at cryogenic temperatures, but the native
window on the hermetically-sealed unit is opaque beyond
1.0~$\mu$m. The SIRMOS team is developing strategies to either replace
the visible window with one transparent to NIR radiation or to operate
without a window. This is a straightforward engineering study that
presents low technical risk.

While the DMD is the preferred multi-object spectroscopy solution for SIRMOS 
due to the superior multiplexing capability, the Next Generation Mechanical Micro-Shutter 
Array (NGMSA) would also work.\cite{10.1117/12.734152}  
Robert Content's accompanying poster on the SIRMOS optical design at this conference describes this alternative 
implementation using the NGMSA, which promises higher throughput at the cost of 
reduced observing efficiency and lower current maturity compared to the DMD option.

The spectrometer's continuous wavelength coverage of 1.25--2.5~$\mu$m
(only possible from space) ensures that SIRMOS captures key galaxy
emission lines useful as redshift diagnostics for all galaxies in the
redshift range $1\le z\le4$ with no redshift gaps. The resolving power
$R=\lambda/\Delta\lambda$ of 1300 enables the clean separation of
\Nii\ and \Ha\ for unambiguous redshift determination.

SIRMOS captures spectra with a Teledyne HAWAII-4RG\tm\ 4K$\times$4K 
(H4RG) HgCdTe focal plane array (FPA).\cite{10.1117/12.926750} This FPA has
space heritage on the \textit{Roman} Wide Field Imager
(WFI) and is available for SIRMOS by special arrangement with
NASA.\cite{10.1117/1.JATIS.6.4.046001} The FPA features a large
detector area with $10~\mu$m pixels, 
outstanding quantum efficiency ($>80$\% on average),
low dark current (about 0.005~$\electron$/s/pix) and low read noise
(5~$\electron$) through multiple non-destructive samples. 
SIRMOS employs the H4RG device's ability to perform multiple non-destructive
reads during a single exposure to further reduce read noise, mitigate
cosmic rays, and recover data from saturated pixels using Up-the-Ramp
(UTR) sampling strategies adopted by JWST and
WFI.\cite{10.1117/12.2060114,10.1117/12.789976} The low cadence of
exposures with SIRMOS (250~s integration time) permits real-time
processing of UTR-sampled images in the on-board detector controller
with only modest demands on the payload processor, vastly reducing the
amount of data transmitted to the ground.\cite{2005PASP..117...94O}

The baseline H4RG FPA is well suited to the mission by virtue of its
efficient power and noise characteristics. This device itself draws
only 32~mW during operation and the corresponding 
Teledyne SIDECAR\tm\ controller
draws 880~mW but can be mounted in a warmer environment to manage the
SIRMOS thermal load. SIRMOS is equipped with a dedicated radiator that
cools the FPAs to 95~K, reducing dark current to deliver higher $S/N$
for the faintest targets. At the assumed limiting magnitude of $K=18$ on the Vega system, the $0.28~\textrm{m}^2$ collecting area of the SIRMOS instrument will deliver about $0.005~\textrm{photons}~\textrm{s}^{-1}~\textrm{px}^{-1}$ which yields about $4.5~\textrm{e}^{-}~\textrm{px}^{-1}$ in each $4\times250$~s exposure sequence. This magnitude limit corresponds to a target density of around 10,000 galaxies deg$^{-2}$, ideal for the sampling density of the MOS. 

The resulting spectrometer collects up to 4000 spectra per exposure for
highly efficient survey operations. At the SIRMOS spectral dispersion,
individual spectra over the 1.25--2.5~$\mu$m range occupy 1785
detector pixels, with the sampling of 2 pixels per spectroscopic
resolution element. The $4\textrm{K}\times4\textrm{K}$ FPA (featuring $4088\times4088$
active pixels) easily accommodates 2 
spectra along the dispersion axis.
Furthermore, a minimum 1-pixel offset of spectra in the cross-dispersion axis, prior knowledge of approximate target redshifts from \textit{Euclid} data, and the low level of continuum flux in the emission line spectra resulting from the $R=1300$ resolving power of the spectrograph allows recovery of accurate redshifts even in the presence of some overlap between adjacent spectra. The resulting high degree of multiplexing enables collection of $10^8$ redshifts over the 3-year SIRMOS mission lifetime.

\subsubsection{Imager Requirements \& Estimates }

The SIRMOS imager's primary function is to sense the location of
pre-selected alignment stars used to confirm the telescope pointing
and sky position angle, thereby ensuring that the defined DMD slit
pattern for a given field aligns with the sky positions of the
spectroscopic targets. The SIRMOS imager is fed from an off-axis
pickoff mirror beside the DMD and thus views a region of sky which is
close to but not overlapping with the spectrometer field of view. The
imager's wavelength coverage is dictated by the wavelength response of
the detector, which can be either visible or NIR. 

The imager can operate effectively 
across a range of FoV and sampling; expanding the FoV provides both more stars to 
sample and a longer ``lever arm'' for more accurate position angle measurements, 
but at the cost of less accurate centroids derived from coarser sampling of the 
individual stellar profiles. The baseline design
adopts a $28\arcmin$ FoV with $0\farcsec83$ pixel sampling, which is sufficient to measure 
pointing to better than $0\farcsec1$ and the sky position angle to
10\arcsec. SIRMOS uses a Teledyne H2RG detector in the imager to meet these performance
requirements.

\subsection{Payload Technical Details}
\subsubsection{Thermal design}

SIRMOS leverages the successful thermal design of SPHEREx to achieve
stable cryogenic temperatures with ample margin. As illustrated below
in Fig.~\ref{fig:SIRMOS_thermal}, the fully passive thermal design
uses three innovative V-groove radiators pioneered by JPL.\cite{Moore_2024} 
The photon shields and
V-grooves together protect the optical payload from spacecraft and
environmental loads while directing thermal radiation to deep space
through low emissivity, specular surfaces at a specifically designed
angle. The optical payload is controlled to two separate temperatures
with separate radiators: whereas SPHEREx operates the FPAs at 55~K and
optics at 80~K to lower self-emission in MWIR, SIRMOS maintains
significantly warmer temperatures of 95~K for the FPAs and 140~K for
the optics operating at SWIR wavelengths. The dedicated FPA radiator
emits heat directly to space to lower the temperature. The telescope
and imager use the telescope body as a radiator, emitting directly to
deep space via reflection off the innermost photon
shield. Strategically placed trim heaters maintain thermal balance at
the desired operating temperature even when the space vehicle
experiences the seasonal eclipses typical of sun-sync low-earth
orbits. The SIRMOS thermal design easily meets the payload's needs.

\begin{figure}[tbp]
    \centering
    \includegraphics[width=0.8\linewidth]{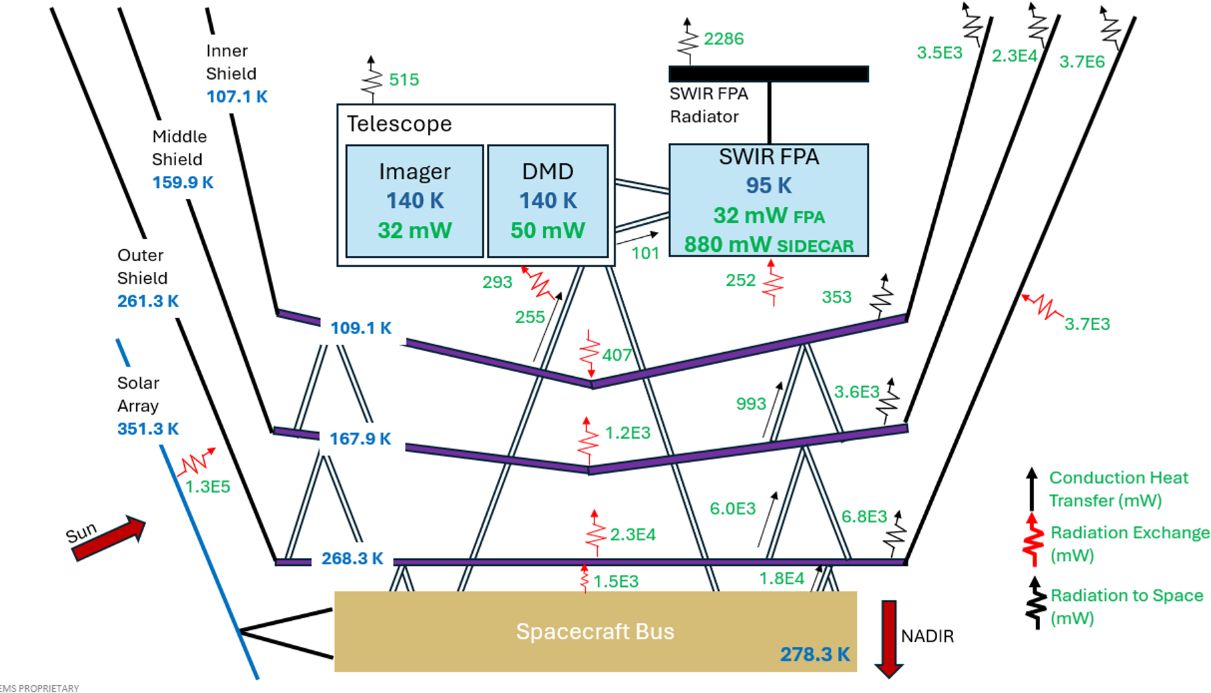}
    \caption{SIRMOS adopts the SPHEREx triple-cone radiator design to cool optics to 140~K and the FPAs to 95~K.}
    \label{fig:SIRMOS_thermal}
\end{figure}

\subsubsection{Mass and dimensions}

\begin{figure}[tbp]
    \centering
    \includegraphics[width=0.75\linewidth]{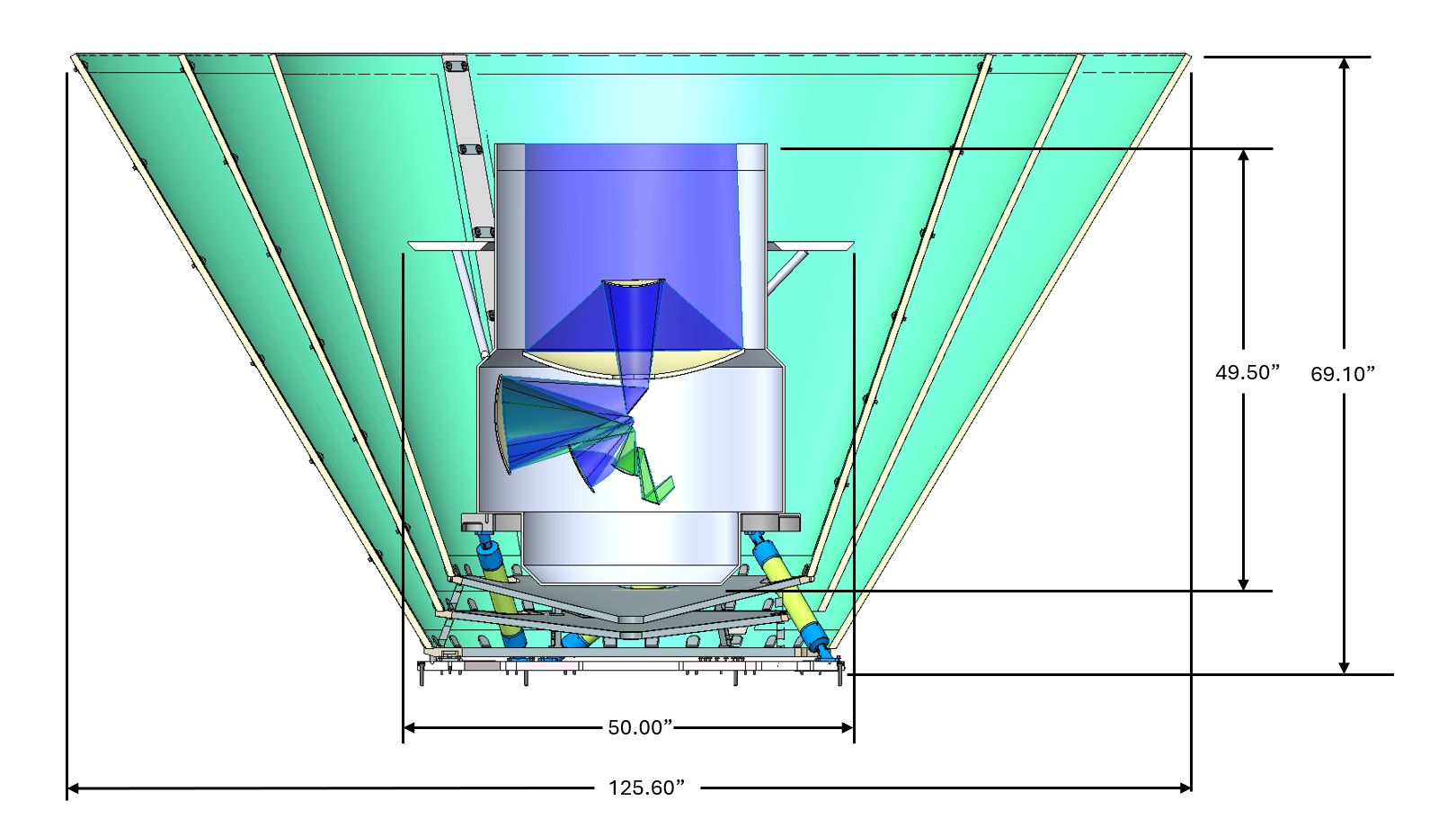}
    \caption{The preliminary SIRMOS instrument design is well sized to fit within the SPHEREx photon shields. The diagram shows the telscope and imager but omits the spectrograph arm for clarity.}
    \label{fig:sirmosVolume}
\end{figure}

SIRMOS has a current best estimate (CBE) payload mass of 116.6~kg; when including
contingency, the total is 146.3~kg. Table~\ref{tab:sirmosMass} breaks
down mass by subsystem. These values include all optics, mounts,
electronics, heaters, radiators, and temperature sensors needed to
operate the payload, including the V-groove radiators and photon
shields.  As shown in Fig.~\ref{fig:sirmosVolume}, the preliminary 
SIRMOS payload volume is sized to fit within the SPHEREx photon shields,
although the exact orientation of the spectrograph arm remains to be defined.
Table~\ref{tab:sirmosParameters} gives the dimensions of both the
inner portion of the payload (excluding the V-groove radiators, light
cones, and bipods) and the full payload including those thermal and
structural elements. A SPHEREx-like bus can readily accommodate the
SIRMOS payload mass and volume.

\begin{table}[h!]
    \centering
    \includegraphics[width=0.65\linewidth]{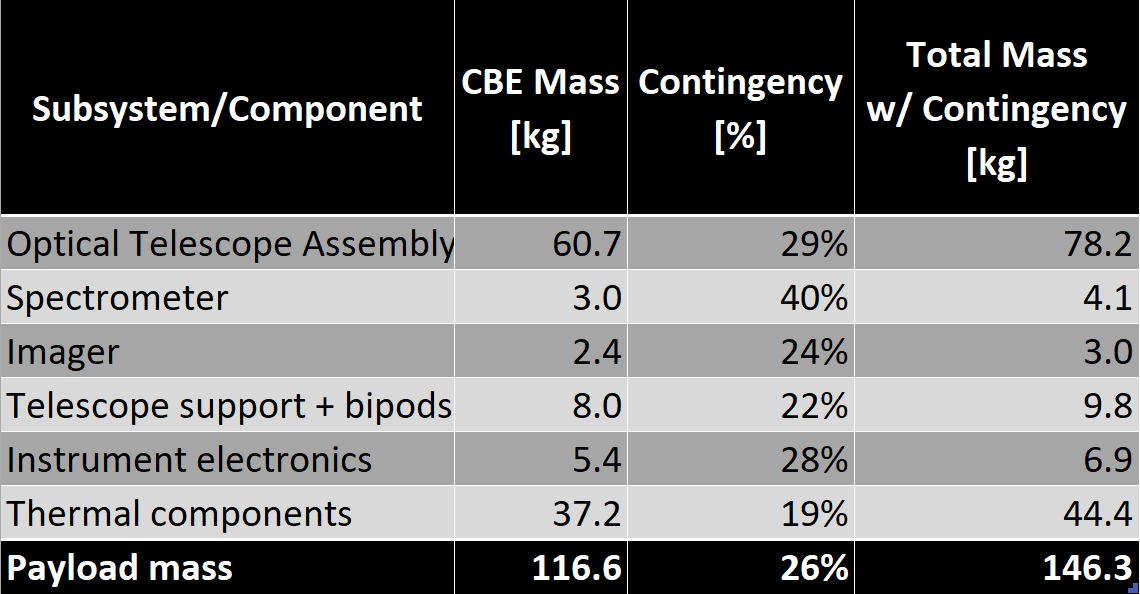}
    \caption{SIRMOS payload mass properties}
    \label{tab:sirmosMass}
\end{table}

\subsubsection{Power requirements}
The SIRMOS payload has relatively low power needs that are well within
the capabilities of the SPHEREx bus. As shown in
Table~\ref{tab:sirmosPower}, the DMD and Focal Plane Module (FPM) are
low-power devices ideally suited for space missions. The payload
electronics power and heater draws are similar to those on
SPHEREx. While its payload nominally operates at a higher temperature
than that of SPHEREx, SIRMOS can run colder if desired to limit the
power drawn by the trim heaters. The SIRMOS payload presents no power
concerns.

\begin{table}[h!]
    \centering
    \includegraphics[width=0.75\linewidth]{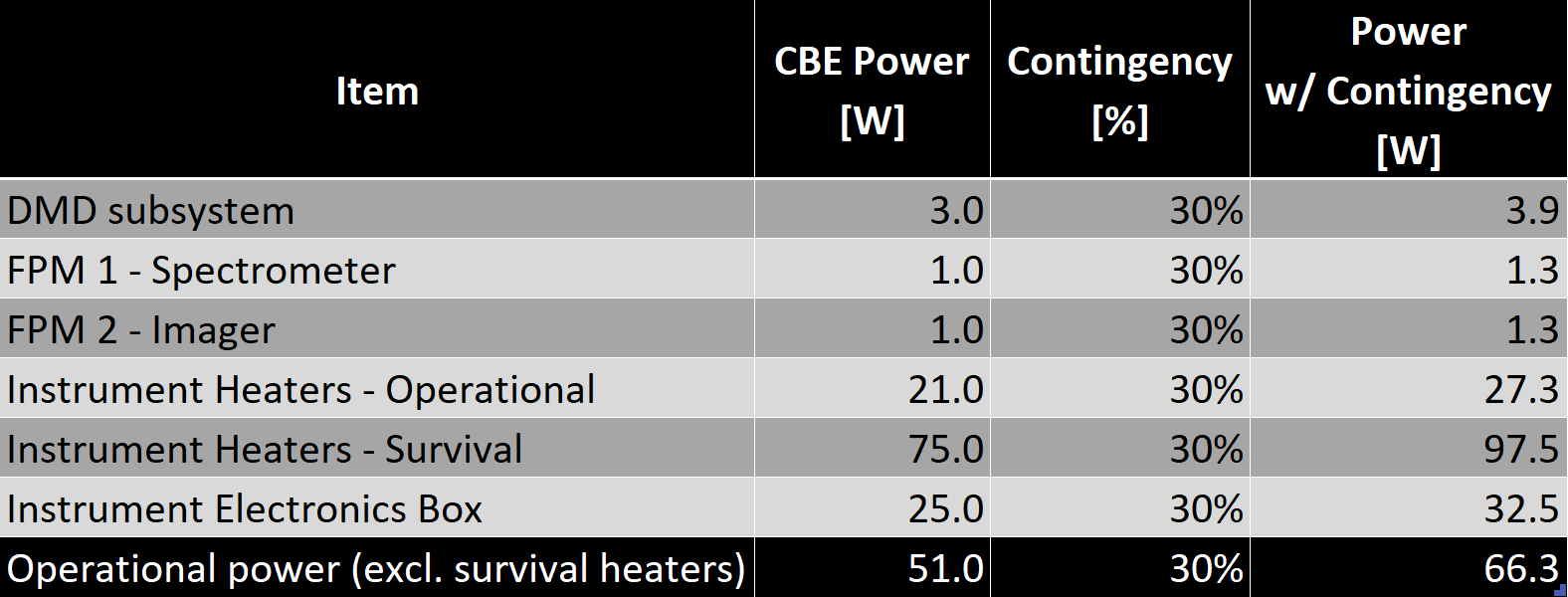}
    \caption{SIRMOS payload power draw of 66~W is comparable to
      SPHEREx and easily accommodated by the same bus} 
    \label{tab:sirmosPower}
\end{table}

\subsubsection{Telemetry requirements}

SIRMOS has modest telemetry downlink needs relative to SPHEREx. SIRMOS
observes each field for 1000~s, acquiring four spectral images each
lasting 250~s. With 20\% overhead, this sequence takes 1200~s per
field; thus, SIRMOS observes a maximum of 72 fields per day and
acquires 288 spectral images. At 16 bits per pixel, each
$4096\times4096$ spectral image occupies 33~MB for a total of
9.7~GB/day. The SIRMOS imager takes periodic images during the
spectral sequence and saves a single 32-bit co-added
$4096\times4096$ image per field, totaling 72 direct images and
another 4.8~GB/day. Assuming a reasonable 50\% on-board data
compression factor, this translates to a total daily science data
volume of 7.2~GB. Housekeeping data collected at 16~kbps adds
0.17~GB/day to give a total daily data volume transmitted to ground of
7.4~GB ($7.4\times10^6$~kB), well within the 97~GB daily data volume
SPHEREx transmits using its Ka-band antenna.

SIRMOS requires once-daily uplink of command telemetry with
observing plans for the coming days, consisting of the following items
for each field: (a)~field center coordinates and slit position angle,
(b)~exposure parameters including start time, exposure duration, and
exposure count, (c)~the desired on/off state for each micromirror in
the DMD array, and (d)~predicted brightness and pixel coordinates of
alignment stars in the imager. A week-long observing plan amounts to
only 139~MB (139,000~kB) per day, easily handled by a 2,000~kbps
S-band receiver. The SIRMOS data needs are thus well within the
capacity of the spacecraft and a SPHEREx-like space-to-ground
communication system.

\subsubsection{Key areas for future design maturation}

While SIRMOS relies primarily on well-developed technologies for components ranging from the bus (previously used on SPHEREx) to the detector (\textit{Roman} heritage), two items merit further attention and development.

\paragraph{DMD re-windowing and space certification.}

DMDs are a mature and proven technology in cinemas around the world
where they operate highly reliably for digital projectors. In recent
years, multiple ground-based astronomical spectrographs have also
successfully employed DMDs as mechanical slitmasks, where they have
also proven to be highly reliable. Two hurdles remain to using DMDs in
space applications. First, the stock windows supplied with the
hermetically-sealed CINEMA\reg\ DMD modules from the manufacturer contain
an anti-reflection coating that has relatively low transmission in the
IR. Manufacturing new substrates with IR-optimized coatings is
straightforward, but the process of achieving a hermetic seal with the
replacement window is not foolproof and can potentially lead to the
introduction of contaminants that could place the delicate piezo
micromirrors at risk of failure. Nevertheless, the recent successes of other research teams in achieving a near-hermetic seal on replacement DMD windows  demonstrates that this hurdle can be overcome.\cite{10.1117/12.2233341,10.1117/12.2691450}

A second impediment to deploying the DMD in SIRMOS is that the CINEMA\reg\
device is not yet space certified. However, LASP plans to fly the CINEMA\reg\
device on the SUMO rocket-borne mission in 2026, thereby demonstrating compatibility with the space environment and achieving Technology Readiness Level 6.\cite{10.1117/12.3019129}

\paragraph{Spectrometer configuration.}

The current incarnation of the SIRMOS instrument presents challenges to accommodate access for integration, alignment, and servicing of critical components (including the focal plane module) while remaining compact enough to fit withing the same volume as SPHEREx to save on bus design effort.  Future work will explore strategies to reconfigure the optics to balance performance, payload volume, and serviceability.

\section{Science program}\label{sxn:science}

Multi-object spectrometers are among our most powerful tools for
astrophysical exploration, and a space-based MOS opens a wide range of
discovery space that includes investigating
breakthrough cosmological science, exploring Galactic star clusters, and
mapping stellar streams in nearby galaxies. SIRMOS aims to apply the
MOS technique to address key cosmological questions by extending
previous efforts to map the large-scale structure of the Universe to
longer wavelengths in the near-IR.
The SIRMOS cosmology survey revisits the same 14,000~deg$^2$ area each year to ensure optimal observations but only takes 60\% of the 3 year prime mission. As shown in Table~\ref{tab:sirmosSurveyParameters}, plans include a generous allotment of time to address secondary science goals and complete community surveys.

\begin{table}[hb]
    \centering
    \includegraphics[width=0.75\linewidth]{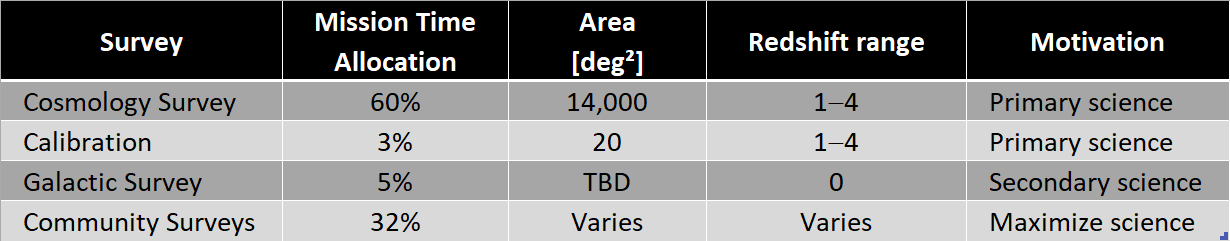}
    \caption{The 3-year SIRMOS mission observation plan features a mix of programs including a significant fraction of time allocated to community science.}
    \label{tab:sirmosSurveyParameters}
\end{table}

Recent years have seen impressive progress in understanding the early
evolution of the Universe in the moments after the Big Bang as well as
the discovery of cosmic acceleration (a.k.a. dark energy).\cite{2020A&A...641A...9P,RevModPhys.75.559}  Much
evidence driving these discoveries comes from observing spatial
fluctuations in the cosmic microwave background (CMB) and from
observations of distant Type~Ia supernovae.\cite{Riess_1998,1999ApJ...517..565P}  However, the distribution
of galaxies captures information imprinted upon the distribution of
matter during the earliest moments of the Big Bang and can be used to
recover crucial evidence revealing the physics
driving cosmological inflation, determining the nature of cosmic acceleration,
and measuring the sum of neutrino masses.  Substantial progress in
this area comes from large galaxy redshift surveys completed using
ground-based MOS (SDSS, DESI) observing in visible light; however,
such surveys are limited in redshift coverage to $z\lesssim2$. A
highly-multiplexed space-based infrared MOS will revolutionize
cosmology by sampling galaxies over the critical $1<z<4$ redshift
range which provides ideal leverage to reduce the existing
uncertainties in cosmological measurements.

\subsection{Probing the origins of the hot Big Bang}

The main science goal for SIRMOS is to discover how the large-scale
structure of the Universe arose from the primordial density
fluctuations that formed in the first $10^{-32}$~s after the big
bang. These fluctuations carry critical information about the
prevailing physical processes driving cosmic inflation and their
statistical distribution is consistent with a single Gaussian random
field to first order; however, more such fields could exist and would
have critical importance for physics.  The presence of multiple
interacting fields would produce detectable deviations from Gaussian
correlations in the spatial distribution of the primordial seeds of the
structures that eventually arose.  Surveys of large-scale structure
that detect such departures can shed important light on the nature of
these interacting fields and constrain their origin.

Determining whether a single field or multiple fields operated during
the inflationary era is of special importance.  In the single-field
model, the power spectrum of fluctuations should exhibit a completely
Gaussian spectrum of correlations; the parameter $\fNL$ characterizing
local-type non-Gaussianity would have a value near zero in this
case. In contrast, if the density fluctuations were created outside of
the inflationary era or arose from multiple non-interacting fields,
then the value of $|\fNL|$ should be of order 1 or
greater.\cite{alvarez2014testinginflationlargescale} A survey testing
whether $|\fNL| << 1$ at a high confidence level can potentially
falsify the single-field inflation model and revolutionize our
understanding of the physics of the early
universe.\cite{2021pdaa.book.....N}

\begin{figure}[tbp]
  \centering
  \hspace{\fill}
  \includegraphics[width=0.45\linewidth]{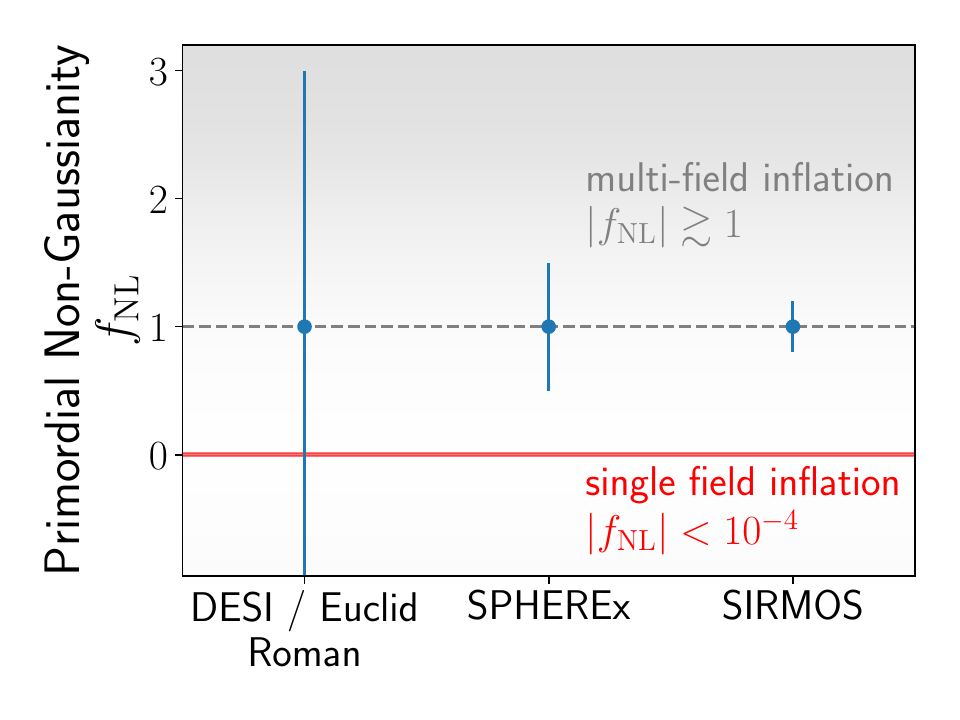}
  \hspace{\fill}
  \includegraphics[width=0.45\linewidth]{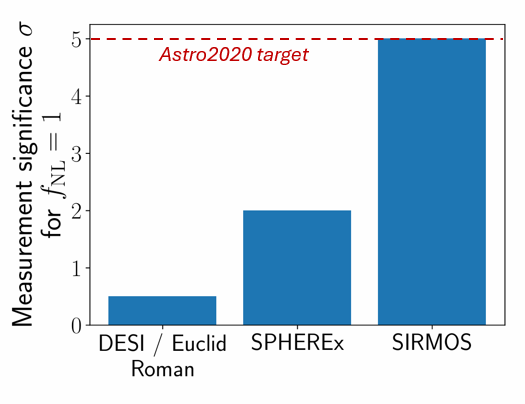}
  \hspace{\fill}
  \label{fig:fNL}
  \caption{\textit{Left:}~Comparison of constraints that various
    surveys place on the non-local Gaussianity parameter
    $\fNL$. \textit{Right:}~Measurement significance that these surveys
    place on $\fNL$ in the $\fNL=1$ case. The ability of SIRMOS
    to survey a large volume at high sampling density with excellent
    redshift accuracy and low systematics enables it to discriminate
    the $\fNL=0$ and $|\fNL|=1$ cases at $5\sigma$ significance. SPHEREx will improve upon the state-of-the-art constraints on $\fNL$ from Planck \cite{2020A&A...641A...9P} by an order of magnitude; SIRMOS will build on SPHEREx heritage to reach the Astro2020 goal for $\fNL$.} 
\end{figure}

To date, the best studies based on the CMB data from Planck have measured
$\fNL=-0.9\pm5.1$ which is too uncertain to discriminate between the
single and multi-field models. Better constraining $\fNL$ is only
possible by employing a three-dimensional survey of galaxies with high
sampling density over a large volume to reduce the impact of cosmic variance,
which is the aim of SIRMOS. By providing continuous wavelength
coverage in the absence of contamination by atmospheric emission
lines, SIRMOS eliminates a critical source of systematic bias and
completes a survey capable of placing unprecedented constraints on
the physics of inflation.
Observing in the near-infrared allows
SIRMOS to sample the full $1<z<4$ redshift range, achieving much
greater redshift sampling depth than existing visible-wavelength
ground-based surveys like DESI and SDSS. The accuracy of the redshifts
from SIRMOS will significantly exceed those of \textit{Euclid},
\textit{Roman}, and SPHEREx, making the survey significantly
  more sensitive to deviations from the standard Gaussian power
  spectrum and thus more sensitive to the presence of multiple
  fields by measuring cosmological parameters precisely to break their degeneracy with $\fNL$. Our analysis indicates that SIRMOS will measure
  $\fNL$ with sufficient accuracy to discriminate between the
  single-field ($\fNL=0$) and multi-field ($|\fNL|\gtrsim1$) models at
  the $5\sigma$ level, outperforming other surveys and achieving the
  ambitious goal set forth in the Astro2020 decadal report
  (Fig.~\ref{fig:fNL}). In summary, SIRMOS is ideally suited to
  provide new evidence to illuminate the origins of the hot Big Bang.

\subsection{Exploring the physics of cosmic acceleration}

The paradigm-changing discovery of cosmic acceleration at the turn of
the 21st century is now well established and implies the existence of
a still-mysterious repulsive gravitational effect ascribed to ``dark
energy'' (DE). The leading theory remains a cosmological constant of
unchanging energy density, but stringent new tests of deviations from
this simple model offer the exciting potential for revealing new
physics or violations of general relativity on the longest
scales. Astronomers are pursuing multiple avenues to study dark
energy, including weak lensing, Type~Ia supernova surveys, galaxy redshift surveys, and galaxy
cluster abundances.
Tantalizing results from
the DESI survey suggest that the DE equation of state may evolve over
time, a result which would invalidate the prevailing cosmological
constant model and require new physics. Existing surveys -- including
DESI -- both lack sensitivity in the $1.5\lesssim z \lesssim4$
redshift range constraining this effect and assume a simplistic linear model
for the time variation of the DE equation of state.

SIRMOS conducts an extensive and accurate survey of galaxies in
this critical redshift gap to provide our most sensitive test yet of
whether dark energy evolves with time, an important probe of its
nature.  With its high-redshift, large-volume, high-accuracy,
well-controlled sample, SIRMOS measures redshift-space distortions
and Baryon Acoustic Oscillations (BAO) to precisely and accurately derive both the cosmic
expansion history and the growth of large scale structure. 
These provide {\it model-independent} constraints on dark energy, to definitively determine whether dark energy is a cosmological constant, a new evolving energy component in the Universe, or the manifestation of gravity requiring the modification of the theory of general relativity.

\subsection{Measuring neutrino masses}

While neutrinos are now known to have non-zero mass, their specific
masses remain poorly constrained, making them the only particles in
the Standard Model to lack known masses. A full understanding of the
physics of our Universe must account for the dynamical contributions
of the neutrino population to the formation of large-scale
structures. A key goal for current physics is to discriminate between
two competing models for the neutrino mass hierarchy: a ``normal''
hierarchy in which one neutrino eigenstate is higher mass and two
others have lower but equal mass, and an ``inverted'' hierarchy
featuring two higher-but-equal-mass states plus a lower-mass third
state. The sum of neutrino masses across the three states differs
significantly in the two models (0.06~eV in the ``normal'' hierarchy
vs. 0.12~eV in the higher one). Because neutrino mass affects the
clustering of galaxies, measurements of large-scale structure in the
galaxy distribution provides critical constraints on neutrino masses.

SIRMOS collects data on the clustering properties of galaxies in
the critical redshift range where predictions of clustering
amplitude derived from perturbation theory benefit from lower
systematic errors. SIRMOS measurements of $1<z<4$ galaxy
clustering amplitude derived from redshift-space distortions can be
compared to CMB-derived measures of the clustering amplitude at
$z=1000$ to confirm or reject the minimum neutrino mass scenario at a
$10\sigma$ level, producing far more stringent constraints than
competing methods can achieve. This is a second area in which SIRMOS
can address a critical problem defined in Astro2020.

\subsection{Secondary science goals}

In addition to probing key questions in cosmology, SIRMOS 
investigates both the
connection between galaxy properties and environment and the properties of our Milky Way galaxy. 
While the primary purpose of the main SIRMOS redshift survey is to employ galaxies as tracers of cosmic large-scale structures, the extensive catalog of emission line galaxy spectra also contains information on the evolving composition and star formation rates of galaxies both before and during the peak era of stellar growth. The SIRMOS catalog  depicts the properties of star forming galaxies across a wide range of environments spanning three orders of magnitude in environmental density, from dense cluster cores to sparse filaments. No existing survey provides spectra of such quality over as large a variety of environments, giving SIRMOS the opportunity to make key contributions to studying galaxy evolution.

While the high degree of dust obscuration in the Milky Way disk prevents
visible-light telescopes from exploring much of our own Galaxy, the IR
spectrometer on SIRMOS opens new volumes to
investigation with unprecedented detail.  A planned survey of the
Galactic plane studies targets across all phases of stellar
evolution ranging from protostars forming deep within dense molecular clouds to
dust-enshrouded Luminous Blue Variables. The high-quality stellar
spectra from SIRMOS will reveal the structure of uncharted regions
within the Milky Way, probe the star formation history of our Galaxy,
and generate a more complete survey of its stellar content.

\section{Conclusions}

The SIRMOS mission is poised to generate breakthrough science addressing key physical questions identified in the Astro2020 decadal survey by providing unparalleled NIR multi-object spectroscopy capability from space. By mapping the large-scale structure of distant galaxies with exceptional breadth and precision, SIRMOS will provide our best constraints yet on cosmic inflation, dark energy, and neutrino masses. Leveraging the expansive \textit{Euclid} imaging covering $\sim$14,000 sq deg, SIRMOS will obtain slit spectra for 100 million galaxies at redshifts from $1<z<4$ using its powerful multi-object spectroscopy capability enabled by a 2048$\times$1080 DMD. SIRMOS uniquely exploits \textit{Euclid}’s deep near-infrared catalog for efficient target selection, collecting high-precision spectra ($R=1300$) with continuous wavelength coverage from 1.25--2.5~$\mu$m enabled by its location above the terrestrial atmosphere to eliminate gaps in redshift measurements and reduce systematic biases. By leveraging heritage bus and thermal designs in the spacecraft design, SIRMOS offers a cost-effective path to obtaining crucial cosmological measurements for the next NASA astrophysics MIDEX opportunity.

\bibliography{report} % bibliography data in report.bib
\bibliographystyle{spiebib} % makes bibtex use spiebib.bst

\end{document}